\documentclass[12pt]{iopart}
\usepackage{graphics}
\usepackage{latexsym}
\newcommand{\bea}{\begin{eqnarray}}
\newcommand{\eea}{\end{eqnarray}}
\newcommand{\bef}{\begin{figure}}
\newcommand{\enf}{\end{figure}}
\newcommand{\ball}{\begin{array}{ll}}
\newcommand{\bal}{\begin{array}{l}}
\newcommand{\ea}{\end{array}}
\newcommand{\xin}{{\xi_i^\nu}}
\newcommand{\xim}{{\xi_i^\mu}}
\newcommand{\xjn}{{\xi_j^\nu}}
\newcommand{\xjm}{{\xi_j^\mu}}
\newcommand{\mup}{{m_\uparrow}}
\newcommand{\mdo}{{m_\downarrow}}
\newcommand{\dmup}{{\Delta m_\uparrow}}
\newcommand{\dmdo}{{\Delta m_\downarrow}}
\newcommand{\aup}{{\alpha_\uparrow}}
\newcommand{\ado}{{\alpha_\downarrow}}
\newcommand{\cupp}{c_\uparrow}
\newcommand{\cdoo}{c_\downarrow}
\newcommand{\eup}{{\mu_\uparrow}}
\newcommand{\edo}{{\mu_\downarrow}}
\newcommand{\rup}{{\rho_\uparrow}}
\newcommand{\rdo}{{\rho_\downarrow}}

\newcommand{\sj}{\sum_{j=1}^N}
\newcommand{\la}{{\langle}}
\newcommand{\ra}{{\rangle}}
\newcommand{\ha}{{1\over 2}}

\begin{document}
\jl{1}
\title[Dynamical properties of a neural network]{Dynamical properties of a randomly diluted neural network with variable activity}
\author{S Gro\ss kinsky\footnote{Present address: Institut f\"ur Angewandte Mathematik, Julius-Maximilians-Universit\"at W\"urzburg, Am Hubland, D-97074 W\"urzburg, Germany} \footnote{E-mail: sngrossk@cip.physik.uni-wuerzburg.de}}
\address{Institute for Theoretical Physics, State University of New York at Stony Brook, Stony Brook, NY 11794, USA}

\begin{abstract}
The subject of study is a neural network with binary neurons, randomly diluted synapses and variable pattern activity. We look at the system with parallel updating using a probabilistic approach to solve the one step dynamics with one condensed pattern. We derive restrictions on the storage capacity and the mutual information content occuring during the retrieval process. Special focus is on the constraints on the threshold for optimal performance. We also look at the effect of noisy updating, giving a dynamical version of the critical temperature, the corresponding threshold and an approximation for the time evolution for small temperatures. The description is applicable to the whole retrieval process in the limit of strong dilution. The analysis is carried out as exactly as possible and over the whole parameter ranges, generalizing some former results.
\end{abstract}

\submitted
\maketitle

\section{Introduction}
Simple cognitive functions of the brain, like associative recall of memories, have been studied for 15 years by using models of attractor networks. One of these is the Hopfield-Little model \cite{hopfield1}, which was analyzed by means of statistical mechanics \cite{amit1} because of the similarity to spin models. To get a more realistical description of properties in data processing, it was extended to models with variable pattern activity \cite{amit2}. In this context a network with neurons $S_i\in\{0,1\}$, instead of $\{ -1,1\}$ was studied in \cite{tsodyks1,buhmann} and showed enhanced storage capacity, resembling the upper bound obtained by Gardner \cite{liesl}. To account for the low connectivity in the brain, this model was extended to random dilution of synapses and analyzed with a dynamical approach \cite{tsodyks2}. The one step parallel dynamics of a network can be solved exactly, using a probabilistic description \cite{bolle2} with a restriction to the first time step or very high dilution \cite{derrida}, due to feed back loops. More recently this method was used to characterize the influence of the threshold on the retrieval properties \cite{bolle1,kitano}.\\
Following this work, we characterize the state of the network by two overlaps with one condensed pattern. We use the probabilistic description of the time evolution of these overlaps to derive conditions for their improvement on the storage capacity. We also consider the mutual information content \cite{bolle1} and the restriction on the threshold in the $(Q,\alpha )$-plane for optimal performance. The effect of positive temperature is studied, leading to an exact expression for the critical value with the corresponding threshold, and an approximation for the time evolution for small temperatures. The analysis is done as exactly as possible, covering the whole range of the pattern activity and the overlaps from a dynamical point of view. In this way we can rederive some former results by looking at special limits, like small pattern activity and the network near retrieval. The main purpose is to demonstrate the possibilities of this relatively simple approach and therefore we study a simple network with realistic features. The most important considerations are confirmed by simulations.\\
In the next two subsections we will introduce the model and the theoretical description. In chapter 2 we will study restrictions on the pattern loading and the threshold for good retrieval properties at zero temperature. The effect of noisy updating is analyzed in chapter 3 and in section 4 we summarize our results.

  \subsection{The model}
The network consists of $N$ neurons $S_i\in\{0,1\}$, $i=1,\ldots,N$ and the state is described by $\mathbf{S}=(S_1 ,\ldots ,S_N )$. There are $p$ patterns \boldmath $\xi^\nu$\unboldmath , $\nu=1,\ldots,p$, with elements $\xin\in\{ 0,1\}$, $i=1,\ldots,N$. They are stored by using a modified Hebb rule (see \cite{buhmann} and \cite{amit2,tsodyks1}):
\bea
J_{ij}={c_{ij} (1-\delta_{ij})\over N c a (1-a)}\sum_{\nu=1}^p(\xin-a)(\xjn-a)
\eea
The $\xin$ are independent, identically distributed random variables (IIDRV) with the distribution function $P(\xin)=a\delta (\xin-1)+(1-a)\delta (\xin)$, $a\in [0,1]$. Therefore the activity of every pattern
\[a^\nu :=\la\xin \ra_i ={1\over N}\sum_{i=1}^N \xin\quad\in\quad [0,1]\]
is a random variable with mean $a$ and variance ${1\over N}a(1-a)$ for all patterns $\nu\in\{ 1,\ldots ,p\}$. The $c_{ij}\in\{ 0,1\}$ are also IIDRV with the distribution $P(c_{ij})=c\delta (c_{ij}-1)+(1-c)\delta (c_{ij})$, $c\in [0,1]$. Hence the number of connections per neuron $i$ over the number of neurons
\[C^i /N:=\la c_{ij}\ra_j ={1\over N}\sum_{j=1}^N c_{ij}\quad\in\quad [0,1]\]
is again a random variable with mean $c$ and variance ${1\over N}c(1-c)$ for all neurons $i\in\{ 1,\ldots ,N\}$.\\
The normalization factor $(ca(1-a))^{-1}$ in the definition of the synapses turns out to be useful to keep the local field in the same order of magnitude over the whole range of pattern activity and network connectivity. As usual we have $J_{ii}=0$ for all $i\in\{ 1,\ldots ,N\}$, enforced by the factor $(1-\delta_{ij})$.\\
\\
The neurons are updated parallel in discrete time steps $t\in \mathbf{N}$ and a uniform threshold $Q$, which may be time dependent, is subtracted from the local field
\[h_i(t):= \sj J_{ij} S_j(t).\]
We have the usual activation function $g(x,\beta )=(1+e^{-2\beta x})^{-1}$, with the noise parameter $\beta=T^{-1}$ and temperature $T$, giving the update rule:
\bea
P(S_i(t+1)=1) = g(h_i(t)-Q,\beta )\nonumber\\
P(S_i(t+1)=0) = 1-g(h_i(t)-Q,\beta )
\label{equdyn1}
\eea
For the noiseless case the activation function reduces to the Heavyside step function, $g(x,\beta )\rightarrow\Theta (x)$ for $\beta\rightarrow\infty$, and the update rule is:
\bea
S_i(t+1) = \left\{ \ball 1 & \mbox{if $h_i(t)-Q>0$}\\
                         0 & \mbox{if $h_i(t)-Q<0$}
                          \ea \right.
\label{equdyn2}
\eea

  \subsection{Theoretical description}
We describe the network theoretically in the thermodynamic limit $N,p\rightarrow\infty$, $\alpha ={p/(cN)} =\mathit{const.}$, denoted with \textit{Lim}, yielding the following simplifications:
\[\mathit{Lim}\ a^\nu =a\ ,\quad\nu=1,\ldots ,p\qquad\qquad\mathit{Lim}\ C^i /N=c\ ,\quad i=1,\ldots ,N\]
We consider only the case of one condensed pattern $\mu\in\{1,\cdots,p\}$ and characterize the state of the network by two corresponding overlaps (omitting the index $\mu$):
\bea
\mup(t):={1\over a}\,\mathit{Lim}\, \la\xim S_i (t)\ra_i &\quad\in&\quad [0,1]\nonumber\\
\mdo(t):={1\over 1-a}\,\mathit{Lim}\, \la(1-\xim)(1-S_i(t))\ra_i &\quad\in&\quad [0,1]\nonumber
\eea
With the two observables one can easily get the current network activity $A(t)$:
\[A(t):=\mathit{Lim}\, \la S_i(t)\ra_i =a\,\mup(t)+(1-a)(1-\mdo(t))\quad\in\quad [0,1]\]
The average overlaps with a noncondensed pattern $\nu\neq\mu$ are $\mup^\nu (t)=A(t)$ and $\mdo^\nu (t)=1-A(t)$. As pattern $\mu$ is condensed, $\mup$ resp. $\mdo$ have to be substentially bigger than these values and the case $\mup=A$, $\mdo=1-A$, which is equivalent with $\mup+\mdo=1$, corresponds to the failure of retrieval. If the system is in a state with $\mup(t)+\mdo(t)=1$ we have $\mup(t+1)+\mdo(t+1)=1$, which can easily be seen from the evolution equation (\ref{evequ2}) derived later. Hence an uncorrelated state will stay uncorrelated.\\
We will often consider the network in a state where $A(t)=a$. Given $\mup$ this is achieved by setting:
\[\mdo(t)={\mdo}_A(t):=1-{a\over (1-a)}(1-\mup(t))\]
To model the time evolution of the network we calculate the mean and the variance of the local field $h_i(t)$ for $\xim=0$ and $\xim=1$, by splitting it in signal part $h_i^S$ and noise part $h_i^N$ as usual:
\bea
h_i =&\mathit{Lim}\, \la{c_{ij}(1-\delta_{ij})\over N c a (1-a)}(\xim-a)(\xjm -a)S_j \ra_j \, + &\qquad\rightarrow\, h_i^S \nonumber\\
&+\, \mathit{Lim}\,\la\la {c_{ij}(1-\delta_{ij})\over N c a (1-a)}(\xin -a)(\xjn -a)S_j \ra\ra_{j,\nu\neq\mu} &\qquad\rightarrow\, h_i^N \nonumber
\eea
The indication of the time dependence is omitted. Now we average over the random distributions of the patterns \boldmath $\xi^\nu$\unboldmath  and the state of the network $\mathbf{S}$ in the thermodynamic limit, fixing the value of $\xim$. Due to self-averaging the signal part has a concrete value, $\eup$ resp. $\edo$, but the noise part is a gaussian random variable with mean $0$ and variance $\sigma^2$:
\bea
\eup=\mathit{Lim}\, \la\la h_i |\xim=1\ra\ra_{\xjm,j\neq i,\mathbf{S}}&=&(1-a)(\mup+\mdo -1) \nonumber\\
\edo=\mathit{Lim}\, \la\la h_i |\xim=0\ra\ra_{\xjm,j\neq i,\mathbf{S}}&=&-a(\mup+\mdo -1) \nonumber\\
\sigma^2 =\mathit{Lim}\, \la\la {h_i^N}^2 \ra\ra_{\xi^\nu ,\nu\neq\mu ,\mathbf{S}}&=&\alpha A
\label{opequn}
\eea
The two random variables $h_i |_{\xim=0,1}-Q$ have mean $\edo-Q$ resp. $\eup-Q$, standard deviation $\sigma$ and the distribution functions
\bea
\rup(x)={1\over\sigma\sqrt{2\pi}}\exp{-(x-\eup+Q)^2\over 2\sigma^2}\qquad\qquad\rdo(x) \mbox{ resp.}
\label{distequ}
\eea
With the dynamical equations~(\ref{equdyn1}),(\ref{equdyn2}) we can compute $\mup \mbox{ and }\mdo$ for the next time step by averaging the activation function over the distribution of the $h_i -Q$.
\bea
\mup(t+1)&=&\la g(h_i(t)-Q,\beta )\ra_\rup =\int_{-\infty}^\infty g(x,\beta )\rup(x)\, dx \nonumber\\
\mdo(t+1)&=&\la 1-g(h_i(t)-Q,\beta )\ra_\rdo =\int_{-\infty}^\infty (1-g(x,\beta ))\rdo(x)\, dx
\label{evequ2}
\eea
For the noiseless case this simplifies to:
\bea
\mup(t+1)&=&\int_0^\infty \rup(x)\,dx=\ha \left( 1+\mathit{erf}\left( {\eup -Q \over \sqrt{2}\sigma}\right)\right)\nonumber\\
\mdo(t+1)&=&\ha \left( 1+\mathit{erf}\left( {Q-\edo\over \sqrt{2}\sigma}\right)\right)
\label{evequ1}
\eea
\\
In the calculations above we considered the neurons to be IIDRVs with activity $A$. This can be realized in the first time step but after one iteration the neurons may be correlated due to feedback loops. In the limit of strong dilution, so that the neurons do not have common ancestors in former time steps, the description is valid for the whole retrieval process. To achieve this the number of connections at each neuron $C^i$ has to be of order $\ln N$, which includes $c\sim N^{-1}\ln N\rightarrow 0$ for $N\rightarrow\infty$. For further details see \cite{bolle2,derrida} and references within there.\\
There are $1/(aN)$ corrections to $\sigma$ and $\eup$ which become relevant in finite systems with small $a$. Therefore we will only present simulations for relativly high $a$, which is not a limitation of the theory.

\section{Dynamical properties at zero temperature}
In the following we look at the one step dynamics and get conditions on the pattern loading and the threshold for improvement of $\mup$ and $\mdo$ at $T=0$. 
We set $\Delta\mup =\mup(t+1)-\mup(t)>0$ and $\Delta\mdo>0$ in equation~(\ref{evequ1}) and get ($\mathit{inverf}(x)$ is the inverse errorfunction $\mathit{erf}^{(-1)}(x)$):
\bea
{\eup -Q\over\sigma}&>&\cupp\ :=\ \sqrt{2}\, \mathit{inverf}(2\mup -1)\nonumber\\
{Q-\edo\over\sigma}&>&\cdoo\ :=\ \sqrt{2}\, \mathit{inverf}(2\mdo -1)
\label{equcond}
\eea

  \subsection{Critical storage capacity and mutual information}
By adding the two equations~(\ref{equcond}) we get the condition $\eup -\edo >\sigma (\cupp + \cdoo)$ for improvement, which has to be satisfied independent of the choice of $Q$, assuming that it is chosen optimally. With the expressions for $\eup$, $\edo$ and $\sigma$ from equation~(\ref{opequn}) we can solve for the critical value $\alpha_c$ where both improvements are zero:
\bea
\alpha_c&=&{(\mup+\mdo -1)^2 \over (\cupp +\cdoo)^2 A}\qquad\mbox{for}\qquad\mup+\mdo\neq 1\nonumber\\
&=&{1\over 2\pi\mup}\ e^{-\cupp^2 /2}\qquad\quad\,\ \mbox{for}\qquad \mup+\mdo=1
\label{equalphacn}
\eea
From the inequality we get the following conditions for improvement of $\mup$ and $\mdo$:
\bea
\alpha <\alpha_c\quad\mbox{if}\quad\mup+\mdo>1\qquad\qquad\alpha >\alpha_c\quad\mbox{if}\quad\mup+\mdo<1
\label{imprcond}
\eea
This is the case because we have $\cupp+\cdoo<0$ for $\mup+\mdo<1$ and in this region we can have improvement for arbitrary high pattern loadings. So if $\alpha$ is too high the network always develops towards a state uncorrelated with the retrieval pattern, where we have $\mup+\mdo=1$.\\
In contrast to the usual storage capacity in equilibrium, $\alpha_c$ is a dynamical variable, depending on $\mup$ and $\mdo$, as well as on $a$ and $c$.\\
\begin{itemize}
\item \textbf{Dependence on network connectivity}\\
As we defined $\alpha ={p/(cN)}$ we see that the maximal number of storable patterns decreases proportional to the number of connections per neuron, because $\alpha_c$ is independent on $c$. This is in accordance with former studies of random dilution in equilibrium (e.g. \cite{tsodyks2}).
\item \textbf{Dependence on pattern activity}\\
For decreasing $a$ the capacity increases monotonically if $\mup+\mdo>1$ and decreases if $\mup+\mdo<1$. If $\mup$ and $\mdo$ are fixed we get the finite value $\alpha |_{a=0}={(\mup+\mdo-1)^2 \over (\cupp+\cdoo)^2 (1-\mdo)}$. When the activity of the network is equal to the one of the stored patterns we have $A=a$, $\mdo=\mdo_A$ and therefore $\cdoo|_{\mdo_A} =\sqrt{2}\mathit{inverf}[1-{2a\over1-a}(1-\mup)]$. By using the approximation $\mathit{inverf}(x)\approx(-\ln{(1-x)})^{1/2}$ for small $x$ (see \cite{erfc}), we get the leading behaviour of $\alpha_c$ for the limit $a\rightarrow 0$:
\bea
\alpha_c \approx {(\mup +\mdo_A -1)^2 \over (\cdoo|_{\mdo_A})^2 a} \approx {\mup^2 c\over -2a\ln{a}}\qquad\mbox{with}\qquad \mup\ll 1-2a
\eea
This is known from former studies \cite{amit2,tsodyks1,buhmann,tsodyks2} as an approximation to the equilibrium storage capacity for small $a$ and resembles the upper bound by Gardner \cite{liesl}.
\item \textbf{Dependence on state of the network}\\
If one of the two parameters, $\mup$ and $\mdo$, is equal to 1 or 0, $\alpha_c$ is zero, because for $\mup,\mdo \rightarrow 1$ we have $\cupp,\cdoo\rightarrow\infty$. This means we can have perfect retrieval only for a finite number of patterns.\\
\end{itemize}
For very small pattern activities the storage capacity may increase but the number of active neurons and the information represented by a single pattern decreases. Therefore the maximal amount of information storable in the network gives a more sensible characterization of performance. To account for the loss of information due to retrieval errors, i.e. $\mup ,\mdo\neq 1$, one uses the mutual information \cite{bolle1}. It can be defined as the negative logarithm of the conditional probability of choosing a network state with $\mup$ and $\mdo$, given the activity $A$:
\bea
\fl\quad\mathit{Lim}\ {I_m \over N}= -\mathit{Lim}\ {1\over N}\ln\left[\left( \ball aN\\ \mup aN\ea\right)\left( \ball (1-a)N\\ (1-\mdo)(1-a)N\ea\right) \left/ \left( \ball N\\ AN\ea\right)\right.\right]\nonumber\\
\lo= a(\mup\ln\mup +(1-\mup)\ln (1-\mup))-(1-A)\ln (1-A)-\nonumber\\
-A\ln A+(1-a)(\mdo\ln\mdo+(1-\mdo)\ln (1-\mdo))\nonumber
\eea
This is the average amount of information per neuron of the retrieval pattern, that can be obtained from the network in a given state $\mup$, $\mdo$. For the maximal mutual information content $i_m$ per synapse in units of bits we get:
\bea
\fl i_m =\mathit{Lim}\, I_m {p_{max}\over c N^2 \ln 2}={\alpha_c\over c\ln 2}\{ &a\left[ \mup\ln{\mup\over A}+(1-\mup)\ln{1-\mup\over 1-A}\right]\nonumber\\
&+(1-a)\left[ \mdo\ln{\mdo\over 1-A}+(1-\mdo)\ln{1-\mdo\over A}\right]\}
\eea
This gives the maximal amount of information per synapse storable in the network, subject to the constraint of improvement of $\mup$ and $\mdo$. It has more reasonable properties for low pattern activity than $\alpha_c$:\\
\begin{itemize}
\item If the network is in a state uncorrelated with the retrieval pattern, i.e. $\mup+\mdo=1$, it is $i_m |_{\mup+\mdo=1}=0$.
\item We have ${i_m}|_{a=0,1}=0$ for any fixed $\mup$ and $\mdo$ with $A\neq a$, as there is certainly no information stored when all pattern elements have the same value.
\item For $\mdo=\mdo_A$ or $A=a$, using the approximation of $\alpha_c$ for small $a$, we get:
\bea
\fl i_m \approx {\mup^2 \over -2a\ln{a}\ln{2}}(-a\mup\ln{a})\ \longrightarrow\  {\mup^3 \over2\ln{2}}\qquad\mbox{for }a\rightarrow 0,\ \mup\ll 1-2a
\eea
That means if we keep the network activity fixed to $a$ during the retrieval process we can obtain a nonzero amount of information from the network for $a\rightarrow 0$, although the information content of a single pattern vanishes. This result is in accordance with former studies in equilibrium (see e.g.~\cite{buhmann,liesl}).
\end{itemize}
The rederivation of $\alpha_c$ and $i_m$ in the limit $a\rightarrow 0$ shows that the enhanced properties for $a\rightarrow 0$ are only given during the retrieval process if we have $A=a$. Further studies showed that this fixed relation between $\mup$ and $\mdo$ is not optimal for maximizing the storage capacity. The optimal relation can be found, but the resulting capacity is of the same order, as the maximum according to Gardner is already reached.

  \subsection{Involving the threshold \label{invq}}
Now we look at the two conditions~(\ref{equcond}) separately. After inserting $\sigma =\sqrt{\alpha A}$ from~(\ref{opequn}) we can solve for $\aup$ resp. $\ado$ where $\dmup=0$ resp. $\dmdo=0$: 
\bea
\aup\ =\ {(\eup -Q)^2 \over\cupp^2 A}\qquad\qquad
\ado\ =\ {(Q-\edo )^2 \over\cdoo^2 A}
\label{alphaqcond}
\eea
Because the conditions were squared only one branch of the parabolas $\aup(Q)$ and $\ado(Q)$ imposes a condition on $\alpha$, depending on the values of $\mup$ and $\mdo$. For $\mup>0.5$ we have $\cupp>0$ and the condition for improvement of $\mup$ (\ref{equcond}) gives an upper bound on $\alpha$. For $\mup<0.5$ the sign of $\cupp$ is changed and therefore we have a lower bound on $\alpha$ for improvement of $\mup$:
\bea
\fl\mup>0.5\Rightarrow\ball \alpha <\aup\,\mbox{ for }Q<\eup \\ \alpha =0\ \ \mbox{ for }Q\geq\eup \ea\qquad\qquad
\mup<0.5\Rightarrow\ball \alpha >\aup\,\mbox{ for }Q>\eup \\ \alpha \geq 0\ \ \mbox{ for }Q\leq\eup \ea\
\label{aupadocond}
\eea
For $\mup=0.5$ the parabola reduces to a vertical line at $Q=\eup$ and we have improvement if $Q<\eup$ for unlimited pattern loading and no improvement for $Q>\eup$. The conditions for improvement of $\mdo$ can be obtain in the same way.\\
\bef
\includegraphics*[15mm,175mm][160mm,253mm]{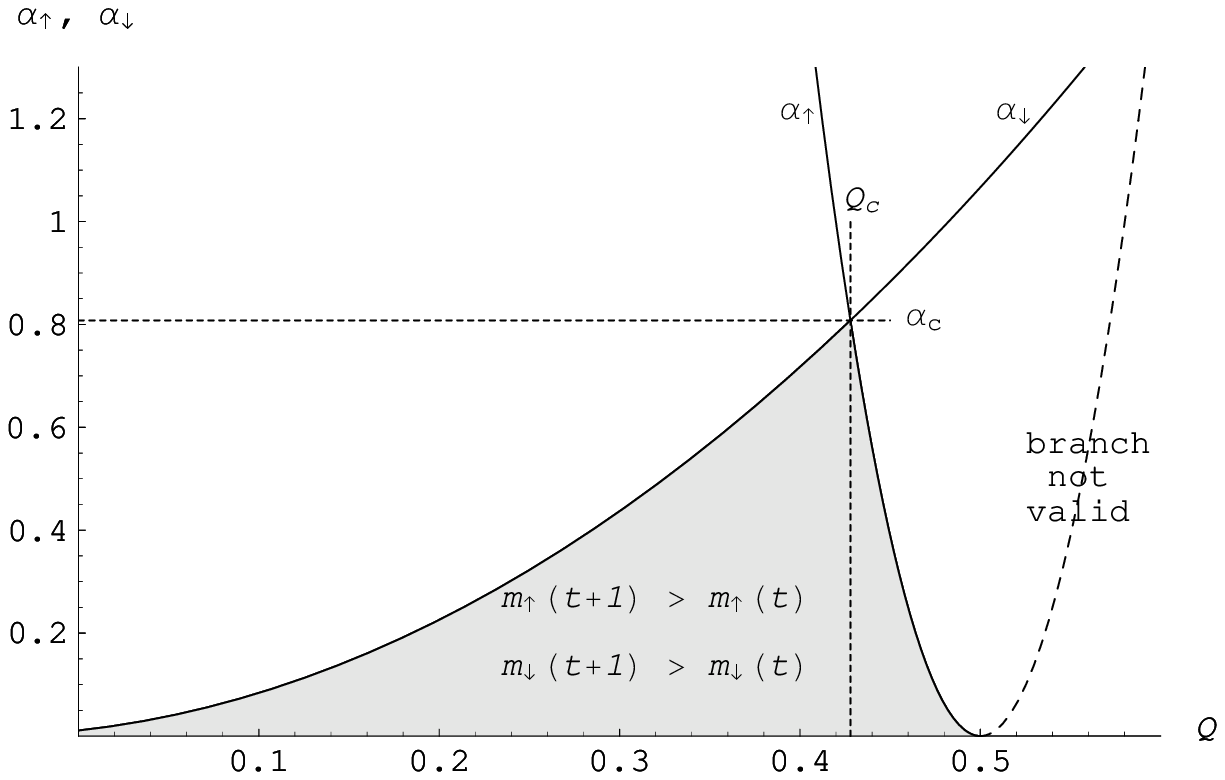}
\caption{$\aup(Q)$ and $\ado(Q)$ for $\mup=0.6$, $A=a=0.1$ and $T=0$. The region of simultaneous improvement is shaded.}
\label{alphaqact}
\enf
For good retrieval qualities we would like to have improvement of both overlaps, which is ensured in a region of the $(Q,\alpha )$-plane limited by the valid branches of $\aup$ and $\ado$ (see figure~\ref{alphaqact}). Hence we have a lower and upper bound for the region of the threshold, $Q\in [\edo +\cdoo\sigma ,\eup -\cupp\sigma ]$ , that will lead to improvements on $\mup$ and $\mdo$ depending on $\alpha$. Unless $\mup=\mdo=0.5$ there is always an intersection of the valid branches of $\aup$ and $\ado$, where both improvements are equal to zero. Here the interval for $Q$ reduces to a single point and we have $\aup =\ado =\alpha_c$. The value of $Q$ at this point is:
\bea
Q_c&=\left( {\cdoo\over\cupp +\cdoo} -a\right) (\mup +\mdo -1)\qquad &\mbox{for}\qquad\mup+\mdo\neq 1\nonumber\\
&={\cdoo\over\sqrt{2\pi}}\ e^{-\cdoo^2 /2}\qquad &\mbox{for}\qquad\mup+\mdo=1
\label{equqcn}
\eea
The dependence on $a$ is linear and for $\mup=\mdo=1$ it reduces to $Q_c =0.5-a$ which is known to be the optimal threshold near retrieval \cite{amitbook,buhmann}. For $\mup+\mdo>1$ the region of simultaneous improvement is bounded below and above and the maximal possible storage is $\alpha_c$. For $\mup+\mdo<1$ the region is only bounded below and $\alpha_c$ is the minimal possible pattern loading for improvement, in accordance with~(\ref{imprcond}). In figure~\ref{alphaqequn} our considerations are confirmed by simulations for $a=0.3$, $\mup=\mdo=0.9$ and $\mup=0.3$, $\mdo=0.9$.\\
\\
\bef
\includegraphics*[15mm,174mm][160mm,253mm]{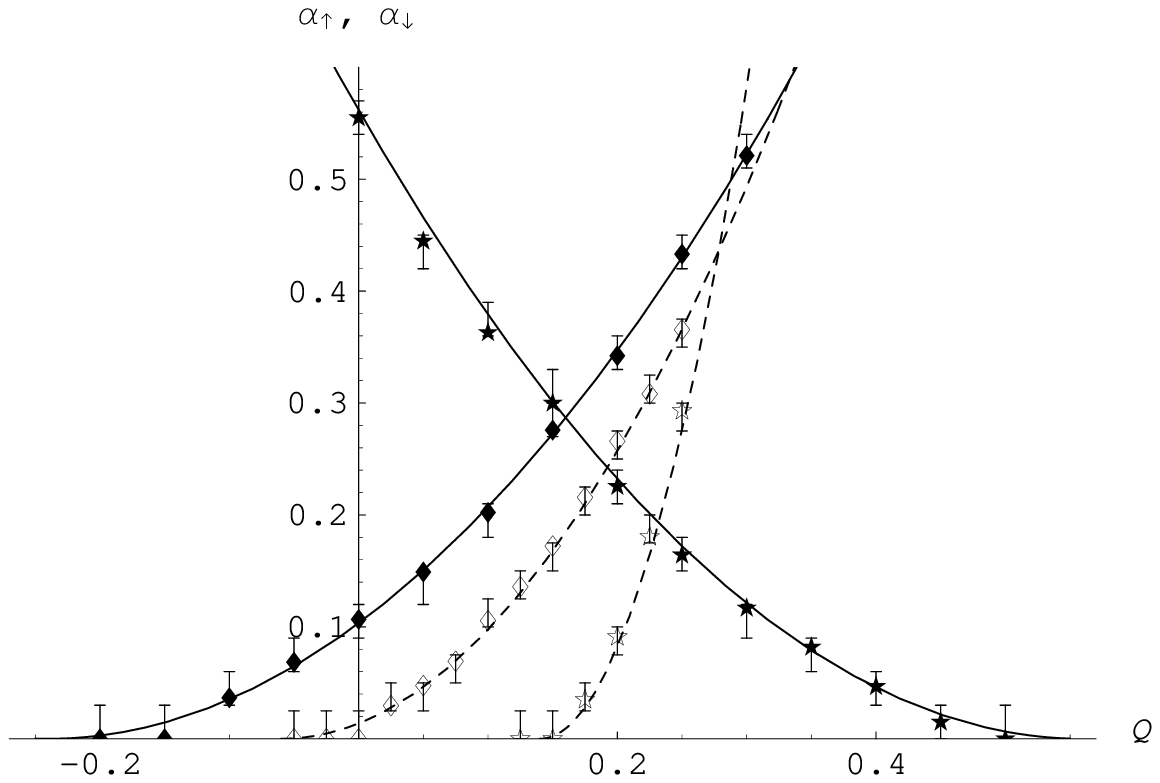}
\caption{Simulations for $\aup(Q)$ and $\ado(Q)$ with $a=0.3$, $T=0$ and $\mup =\mdo =0.9$ \full\  (Data filled), $\mup=0.3,\mdo=0.9$ \broken\  (Data unfilled). Data obtained with $N=600$ and average over 10 networks each with at least 150 stimuli. The average improvement of $\mup$ and $\mdo$ was recorded as a function of $\alpha$ for different values of $Q$. At $\alpha =\aup$ resp. $\ado$, $\dmup$ resp. $\dmdo$ vanishes.}
\label{alphaqequn}
\enf
To illustrate the use of the derived restrictions in the $(Q,\alpha )$-plane we discuss some choices of thresholds as functions of the state of the network which ensure $Q|_{\alpha =\alpha_c}=Q_c$, i.e. they allow for the critical storage capacity $\alpha_c$ (see figure~\ref{qoptqn}).\\
\begin{itemize}
\item \textbf{Critical threshold}\\
The easiest thing is of course to choose $Q=Q_c$ which certainly meets the requirement. But if $\mup\mbox{ resp. }\mdo<0.5$ it does not improve if $\alpha <\alpha_c$ (see $\mup$ in figure~\ref{alphaqequn}). This choice of $Q$ maximizes the term $\Delta\cupp +\Delta\cdoo$.
\item \textbf{Maximizing $\mathbf{\dmup +\dmdo}$}\\
Just for comparison we will also look at $Q=Q_m:={1\over 2}(\eup +\edo)$ which maximizes $\dmup +\dmdo$. This seems to be a somehow `natural' choice, but it only fulfills $Q_m |_{\alpha =\alpha_c}=Q_c$ when $\mup=\mdo$, where we have $Q_m =Q_c$.
\item \textbf{Ensuring $\mathbf{A(t+1)=a}$}\\
Another possibility is to choose $Q=Q_a$ in order to get the network activity $A(t+1)=a$, ensuring an enhanced storage capacity for small $a$. In the retrieval state we also have $A=a$, so we can reach it with this threshold dependence. However the condition $Q_a |_{\alpha =\alpha_c}=Q_c$ is only obeyed for $A(t)=a$.
\item \textbf{Preserving $\mathbf{\mup/\mdo}$}\\
We can also choose $Q=Q_r$ to preserve the ratio $\mup (t+1)/\mdo (t+1)=\mup (t)/\mdo (t)=r$. We only look at this choice because it ensures $Q_r |_{\alpha =\alpha_c } =Q_c$, but if we start out with a ratio considerably different from 1 the attractor reached with this threshold won't be very close to the retrieval state. So the only case where this is really interesting is for $r=1$ where we have $Q_r =Q_c =Q_m$.
\end{itemize}
\bef
\includegraphics*[15mm,173mm][160mm,253mm]{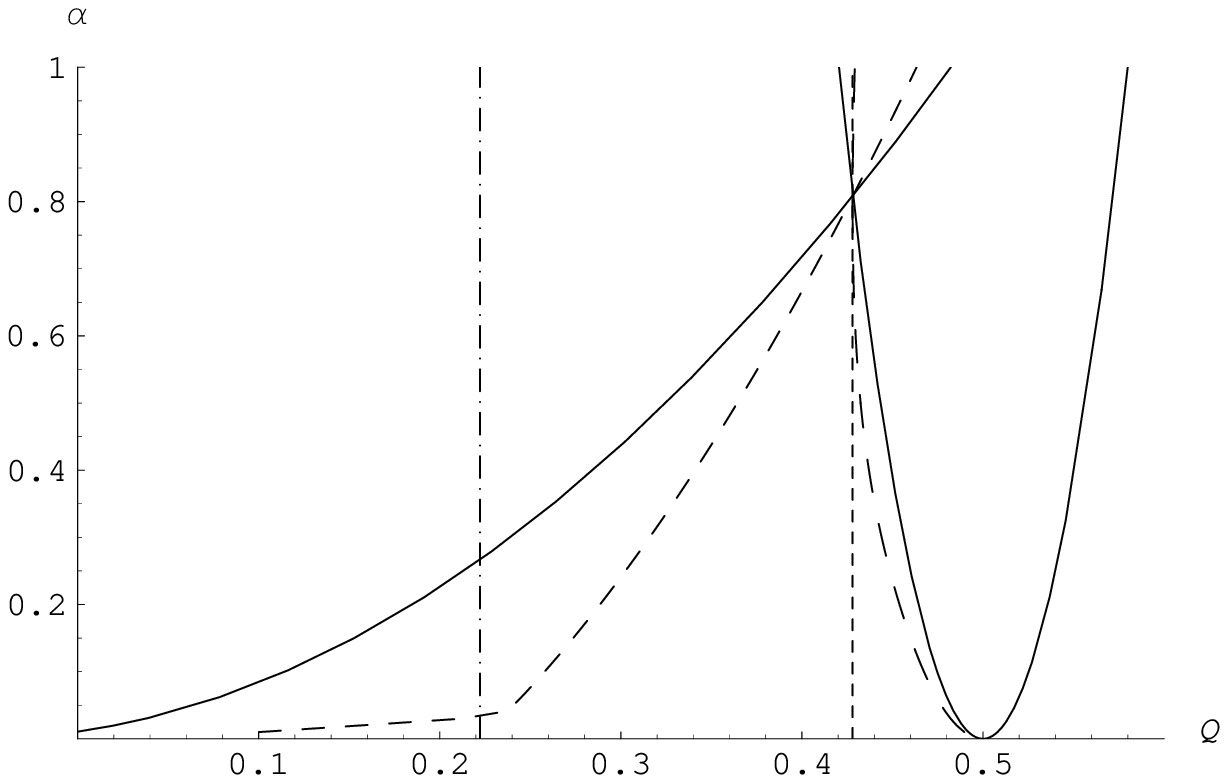}
\caption{Comparison of thresholds for $\mup =0.6$, $A=a=0.1$ and $T=0$. Threshold choices: $Q_c$ \dashed, $Q_m$ \chain, $Q_a$ \broken, $Q_r$ \longbroken.}
\label{qoptqn}
\enf
In figure~\ref{qoptqn} we illustrate the four choices in the $(Q,\alpha )$-plane with $a=0.1$, $\mup=0.6$ and $\mdo=\mdo_A =0.956$. Therefore we have $Q_a |_{\alpha =\alpha_c}=Q_c$ and the curve of $Q_m$ does not cross the critical point $(Q_c ,\alpha_c )$. One can compare the different choices of thresholds by the improvements $\Delta\mup$ and $\Delta\mdo$, which can easily be calculated. As $\mup=0.6$ the upper limit for $\Delta\mup$ is $0.4$ and for $\Delta\mdo$ it is $1-\mdo_A =0.044$. For small $\alpha$ these limits are reached by $Q_c$ and $Q_a$. As expected $Q_a$ gives the highest improvements because it has the best position between the two parabolas in figure~\ref{qoptqn}. The improvement of $\mup$ resp. $\mdo$ is a monotonically increasing function of the distance between $Q$ and $Q|_{\alpha=\aup}$ resp. $Q|_{\alpha=\ado}$. $Q_c$ and $Q_r$ give the highest possible storage capacity $\alpha_c$ in any case but the improvements $\dmup$ look rather poor, especially of $Q_r$.\\
If we consider a case with $\mup=\mdo$ the picture is symmetric with respect to the axis $Q=Q_c$ and therefore $Q_m =Q_r =Q_c$ give the best possible improvements $\dmup=\dmdo$. Thus the optimal choice of the threshold during the retrieval process depends on the situation. If one has $\mup(0)\approx\mdo(0)$ and $a\approx 0.5$ the best choice for fast retrieval is $Q_c$. But in the case of small pattern activity this limits the storage capacity and it is better to choose $Q=Q_a$, independent of the initial condition.\\

\section{Dynamics for positive temperature}
With increasing temperature the critical storage capacity, as defined in $2.1$, decreases and we take $T_c$ as the value of $T$ where $\alpha_c =0$. The allowed region for the threshold reduces to a single point $Q_c |_{T=T_c}$, which we will also calculate. After that we make an approximation for $\aup$ and $\ado$ for small temperatures.

  \subsection{Critical temperature}
We look at the dynamical equation~(\ref{evequ2}) in the limit $\alpha\rightarrow 0$, where the distribution functions $\rup$ resp. $\rdo$ reduce to delta functions, because $\sigma^2 \sim\alpha$:
\bea
\fl\rup(x)={1\over\sigma\sqrt{2\pi}}\exp{-(x-\eup+Q)^2\over 2\sigma^2}\ \stackrel{\alpha\rightarrow 0}{\longrightarrow}\ \delta (x-(\eup -Q))\qquad\qquad\rdo(x)\mbox{ resp.}\nonumber
\eea
In this limit we can evaluate the integrals in the evolution equation~(\ref{evequ2}) and after solving for $\beta$ and $Q$ we get the critical temperature $T_c =\beta_c^{-1}$ and the corresponding threshold $Q_c |_{T=T_c}$.\\
For $\mup+\mdo\neq 1$:
\bea
T_c &=&-2(\mup +\mdo -1)\over \ln ({1\over\mdo}-1) +\ln ({1\over\mup}-1) \\
Q_c |_{T=T_c}&=&\left({\ln ({1\over\mdo}-1)\over\ln ({1\over\mdo}-1) +\ln ({1\over\mup}-1)}-a\right) (\mup+\mdo-1)
\label{tcequn}
\eea
For $\mup+\mdo=1$:
\bea
T_c\ =\ 2\mup(1-\mup)\qquad\qquad Q_c |_{T=T_c}\ =\ \mup(1-\mup)\ln \left({1\over\mup}-1\right)\nonumber
\eea
The critical temperature is independent of the pattern activity and the network connectivity. It is easy to see that the maximum value is $T_c =0.5$ at $\mup=\mdo=0.5$, and if at least one of the two observables is equal to zero or one we have $T_c =0$. We see $T_c (\mdo)$ in figure~\ref{tcn} for several values of $\mup$, confirmed by simulations which were done at different values of $a$ and $c$ to demonstrate that it is independent on these parameters.\\
\bef
\includegraphics*[15mm,173mm][160mm,255mm]{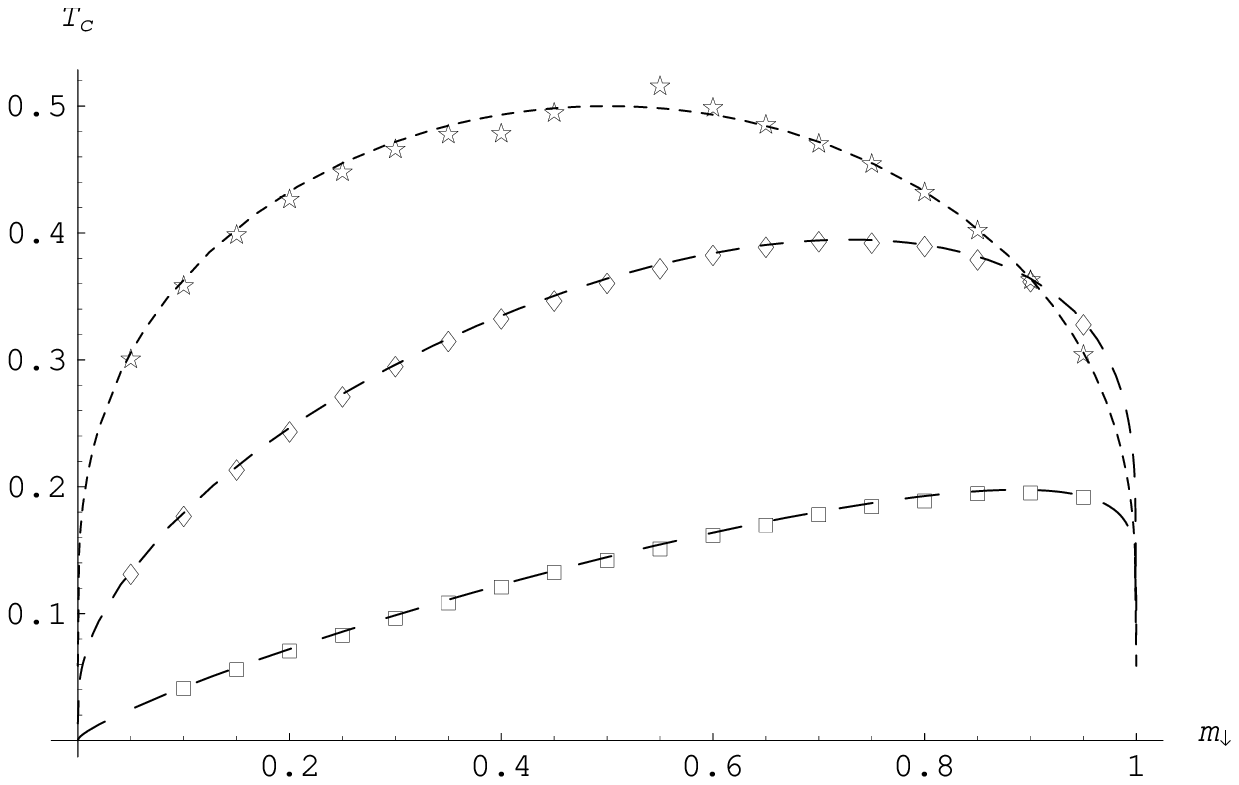}
\caption{Critical temperature $T_c (\mdo)$ for $\mup =0.5$ \dashed, $\mup =0.9$ \broken, $\mup =0.999$ \longbroken. Data with $N=2000$, $\alpha =0.0005$ and $\mup=0.5$ ($\star$, $a=0.3$ and $c=0.5$), $\mup=0.9$ ($\opendiamond$, $a=0.3$ and $c=1$) and $\mup=0.999$ ($\opensquare$, $a=0.5$ and $c=1$).}
\label{tcn}
\enf
The critical threshold at $T=T_c$ has the same linear dependence on $a$ as for zero temperature, except for an additive constant depending on $\mup$ and $\mdo$. Only for $\mup,\mdo\approx 1$ with $\mup\neq\mdo$ there is a considerable difference between the two, in the case $\mup=\mdo$ both are equal for every $a$. This indicates that for increasing temperature the critical point in the $(Q,\alpha )$-plane is mainly shifted to lower $\alpha$ with fixed $Q$, which is true to high accuracy as we will see next.

  \subsection{Expansion for small temperatures}
For $T>0$ we solve the conditions $\mup (t+1)=\mup (t)$ and $\mdo (t+1)=\mdo (t)$ for $\aup$ and $\ado$ by using an approximation for low temperatures, as $T_c \in [0,0.5]$. In the following we look at the first condition in order to get $\aup$, the same can be done for $\ado$. After some algebra given in the appendix we have the following approximation which is accurate for $T\ll {2\sigma^2 /(\eup-Q)}$:
\bea
\mup(t+1)\, \approx\, \mup(t+1)|_{T=0}-T^2{\eup-Q\over 2\sqrt{2\pi}\sigma^3}\ e^{-({\eup-Q\over\sqrt{2}\sigma})^2}{\pi^2 \over 12}
\eea
By looking at the numerical solutions of $\aup(Q)$ we see that the shape of the parabola is more or less unaltered, it is just shifted down. Therefore we make the ansatz $\aup=\aup|_{T=0}-f(T)$ where the function $f(T)$ accounts for the downshift of $\aup$ as an additional source of noise, like in \cite{amit1}. Now we replace $\aup|_{T=0}$ by $\aup+f(T)$ in the dynamical equation for zero temperature (\ref{evequ1}) and expand at $f(T)=0$ to first order:
\[\mup(t+1)\approx\mup(t+1)|_{T=0}-{\eup-Q\over 2\sqrt{2\pi}\sigma^3}\ e^{-({\eup-Q\over\sqrt{2}\sigma})^2}Af(T)\]
This term looks very similar to the expansion of the dynamical equation and by comparing the two we get the following approximation:
\bea
\aup_1 \approx\aup|_{T=0}-\gamma_1 T^2\qquad\ado_1 \approx\ado|_{T=0}-\gamma_1 T^2 \quad\mbox{where}\ \gamma_1={\pi^2 \over 12 A}
\label{appr1}
\eea
The calculation for $\ado$ gives the same correction to this order. We labeled the approximation with 1 because we can also think of another one. By assuming a quadratic $T$-dependence of $f$ we can make directly the ansatz $\aup=\aup|_{T=0}-\gamma_2 T^2$. Knowing the critical temperature we get the condition $\alpha_c |_{T=T_c}=\alpha_c |_{T=0}-\gamma_2 T_C^2 =0$, yielding another approximation:
\bea
\aup_2 \approx\aup|_{T=0}-\gamma_2 T^2\qquad\qquad\ado_2 \approx\ado|_{T=0}-\gamma_2 T^2\nonumber\\
\mbox{where}\qquad\gamma_2 ={1\over 4A}{\left[\ln{({1\over \mup}-1)}+\ln{({1\over \mdo}-1)}\right]^2\over (\cupp+\cdoo)^2}
\label{appr2}
\eea 
\bef
\includegraphics*[15mm,175mm][160mm,253mm]{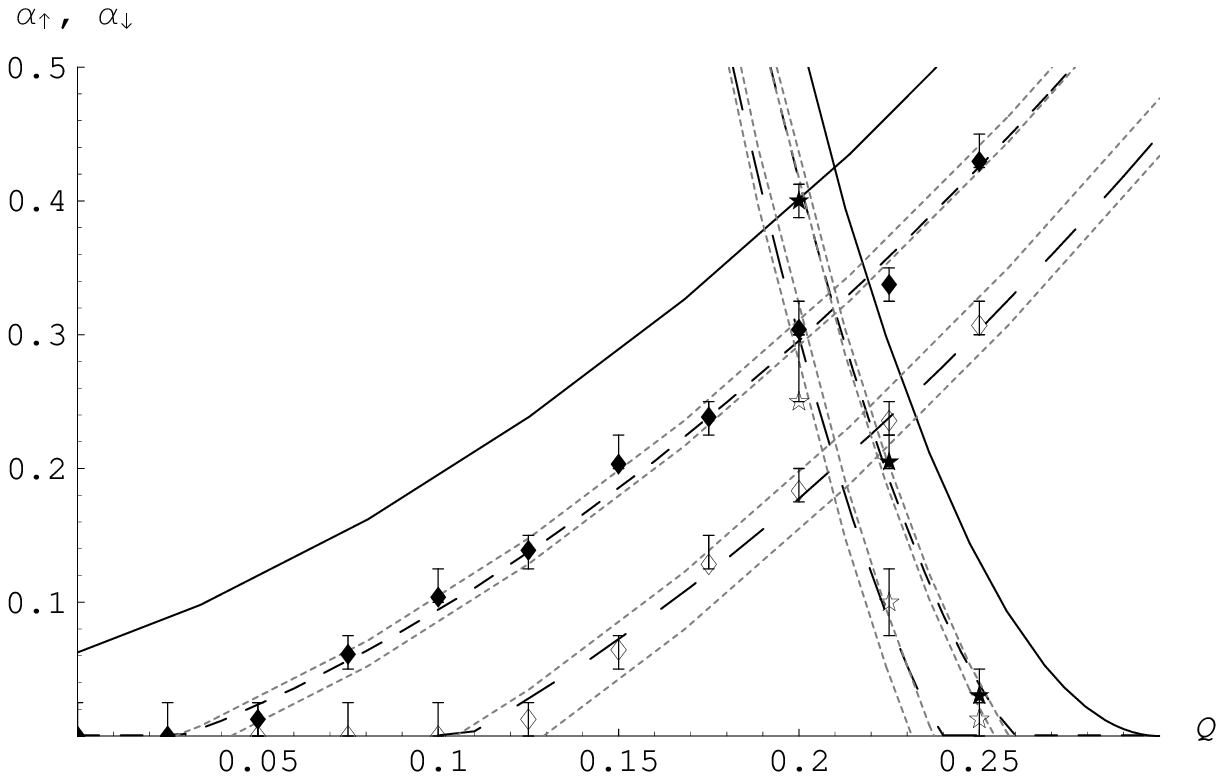}
\caption{$\aup(Q)$ and $\ado(Q)$ for $\mup =0.6$, $A=a=0.3$ and $T=0$ \full, $T=0.2$ \broken\ (Data filled), $T=0.3$ \longbroken\ (Data unfilled). Approximations \full\ (grey lines): $\aup_1$ and $\ado_1$ lower curves, $\aup_2$ and $\ado_2$ upper curves.}
\label{alphaqtn}
\enf
In figure~\ref{alphaqtn} both approximations are compared with the numerical solution for $T=0.2$, $0.3$ and the case $T=0$ for $\mup=0.6$ and $a=A=0.3$. Therefore we have $\gamma_1 =0.822{1\over A} > \gamma_2 =0.679{1\over A}$ and the first approximation~(\ref{appr1}) gives the lower curves, the second~(\ref{appr2}) the upper ones. As expected (\ref{appr2}) is better for small $\alpha$, as $T_c$ was evaluated at $\alpha =0$. For $\aup_1$ and $\ado_1$ to be accurate we have the condition $\alpha\gg \max (\eup-Q,Q-\edo) {1\over 2A}T\approx 0.1$ from above, making it better for higher storage levels.\\
We also see that with increasing temperature the ansatz of a simple downshift of the parabolas $\aup(Q)$ and $\ado(Q)$ becomes more and more inaccurate. But even for $T=0.3$, which is relatively high, the approximations are pretty good. Therefore $\alpha_c$ decreases proportional to $T^2$ for small temperatures, which is in accordance with former studies in equilibrium (see e.g. \cite{amitbook}). From the $1/A$-dependence of $\gamma_1$ and $\gamma_2$ we know that the effect of positive temperature is much more drastic for small network activities than for large ones.

\section{Conclusion}
In this work we described a randomly diluted neural network model with variable pattern activity using a probabilistic approach to solve the one step dynamics with one condensed pattern. By carrying out the analysis as exactly as possible we were able to confirm with this relatively simple approach many previous results on this model, derived with different techniques and often restricted to special cases. Most of our results are valid for the whole range of the different parameters, except for the dilution level, generalizing some former studies.\\
We got new insight in the dynamical properties of the network, studying the critical storage capacity, information content and critical temperature for arbitrary network states. Special focus was on the resulting constraints on the threshold to realize the critical values, a feature that was often overlooked in former studies. We used this to analyze the effects of choices of threshold functions during retrieval. We also showed that we have to impose conditions during the retrieval process to get the enhanced properties derived in former studies.\\
To demonstrate the possibilities of the probabilistic approach we chose a neural network which is relatively simple, but has many realistic features. The analysis is not restricted to this model, it can be used for all networks  with parallel updating. Further important is the site-independent description of the local field, in our case represented by $\eup$, $\edo$ and $\sigma$, excluding for example site-dependent thresholds. But the method is easily extendable to other models with graded response neurons (see~\cite{bolle2}), groups of patterns with different activities, a finite number of condensed patterns or sequential patterns (see \cite{kitano}).\\
It may be interesting to extend the presented analysis to these cases, making it possible to describe various network properties exactly with relatively easy computations. Especially the restrictions we derived on the parameters during retrieval may be very helpful for simulations.\\

\ack
I am thankful to professor Robert Shorck for advising my work on this subject and to Miguel Maravall for very useful discussions. This work was supported in part by the german academic exchange program (DAAD) and the Swartz Foundation at Stony Brook.

\appendix
\section*{Appendix. Calculation for $T>0$}
In section 3.2 we want to calculate $\aup$ and $\ado$ from the dynamical equations for $T>0$. In the following we perform the calculation for $\aup$, where $\ado$ can be evaluated in exactly the same way. First we split the integral in the dynamical equation~(\ref{evequ2}):
\[\mup(t+1)=\int_{-\infty}^0 {\exp (2\beta x)\over 1+\exp (2\beta x)}\rup(x)\, dx +\int_0^\infty {1\over 1+\exp (-2\beta x)}\rup(x)\, dx\]
By using the geometric series $1/(1+y)=\sum_{n=0}^\infty (-1)^n y^n$ for $y<1$ and replacing $x$ by $-x$ in the first integral we get:
\[\mup(t+1)=\sum_{n=0}^\infty (-1)^n \int_0^\infty \left[e^{-2\beta xn}\rup(x)-e^{-2\beta x(n+1)}\rup(-x)\right] \, dx\]
Inserting $\rup(x)$ from equation~(\ref{distequ}), we have to calculate two gaussian integrals of the form
\[\int_0^\infty e^{-q_1 x-q_2 x^2}\, dx={\sqrt{\pi}\over 2\sqrt{q_2}}\exp\left( {q_1^2\over 4q_2}\right)\left[ 1-\mathit{erf}\left( {q_1 \over2\sqrt{q_2}}\right) \right].\]
After this we see that $\mup(t+1)|_{T=0}$ is the term in the sum for $n=0$ and we get the expression
\[\mup(t+1)=\ha\left(1+\mathit{erf}\left( {\eup-Q\over \sqrt{2}\sigma}\right) \right) -\Delta =\mup(t+1)|_{T=0}-\Delta\]
with the correction term
\bea
\Delta=\ha\sum_{n=1}^\infty (-1)^n e^{2\beta^2 \sigma^2 n^2}&\left[ e^{2\beta (\eup-Q)n}\left( 1-\mathit{erf}\left( \sqrt{2}\sigma\beta n+{\eup-Q\over\sqrt{2}\sigma}\right) \right) -\right. \nonumber\\
&\left. -e^{-2\beta (\eup-Q)n}\left( 1-\mathit{erf}\left( \sqrt{2}\sigma\beta n-{\eup-Q\over\sqrt{2}\sigma}\right) \right) \right] .\nonumber
\eea
This expression is still exact but in order to do the sum we use the approximation $1-\mathit{erf}(x)=\exp (-x^2)/(\sqrt{\pi}x)$ \cite{erfc}. This is accurate for $x\gg 1$ which means in our case $\beta\gg (\eup-Q)/(2\sigma^2)$. The expression for $\Delta$ simplifies to
\[\Delta={\eup-Q\over 2\sqrt{2\pi}\beta^2 \sigma^3}\ e^{-\left({\eup-Q\over\sqrt{2}\sigma}\right)^2 } \sum_{n=1}^\infty (-1)^n \left( n^2 -\left( {\eup-Q\over 2\beta\sigma^2}\right) ^2 \right) ^{-1}.\]
The sum can be done exactly and we end up with the following, writing $T=\beta^{-1}$:
\bea
\mup(t+1)=\mup(t+1)|_{T=0}-T^2{\eup-Q\over 2\sqrt{2\pi}\sigma^3}\ e^{-({\eup-Q\over\sqrt{2}\sigma})^2}\Gamma(T\pi{\eup-Q\over 2\sigma^2})\nonumber\\
\Gamma(x)={x-\sin{x}\over 2x^2\sin{x}}={\pi^2\over 12}+{7\pi^4\over 720}x^2+\ldots\nonumber
\eea
For $T\ll 2\sigma^2/(\eup-Q)$ the argument of $\Gamma$ is small and we use the zero order approximation $\Gamma\approx {\pi^2\over 12}$.

\end{document}